\documentclass[a4paper,12pt]{article}
\usepackage{amsmath,amssymb, amsthm}

\usepackage{graphicx}
\usepackage{color}
\usepackage{geometry}
\usepackage{psfrag}
\usepackage{epsfig}

\usepackage{a4}
\usepackage{amssymb,latexsym}
\usepackage[mathscr]{eucal}
\usepackage{cite}
\usepackage{url}

\DeclareFontFamily{OT1}{rsfs}{} \DeclareFontShape{OT1}{rsfs}{m}{n}{
<-7> rsfs5 <7-10> rsfs7 <10-> rsfs10}{}
\DeclareMathAlphabet{\mycal}{OT1}{rsfs}{m}{n}

\newcommand{\be}{\begin{equation}}
\newcommand{\ee}{\end{equation}}

\newcommand{\bea}{\begin{eqnarray}}
\newcommand{\beaa}{\begin{eqnarray*}}
\newcommand{\bean}{\begin{eqnarray}\nonumber}

\newcommand{\eea}{\end{eqnarray}}
\newcommand{\eeaa}{\end{eqnarray*}}

\newcommand{\bel}[1]{\begin{equation}\label{#1}}
\newcommand{\beal}[1]{\begin{eqnarray}\label{#1}}
\newcommand{\beadl}[1]{\begin{deqarr}\label{#1}}
\newcommand{\eeadl}[1]{\arrlabel{#1}\end{deqarr}}
\newcommand{\eeal}[1]{\label{#1}\end{eqnarray}}

\newcommand{\tr}{\operatorname{tr}}
\newcommand{\Div}{\operatorname{div}}

\theoremstyle{plain}
\newtheorem{theo}{Th�or�me}[section]
\newtheorem{prop}[theo]{Proposition}

\newtheorem{Lemma}[theo]{Lemma}
\newtheorem{Theorem}[theo]{Theorem}

\theoremstyle{definition}

\theoremstyle{remark}

\newtheorem{engrk}[theo]{Remark}

\def \R {\mathbb{R}}

\newcommand{\mcM}{{\mathcal M}}

\newcommand{\mcU}{{\mathcal U}}

\newcommand{\mcK}{{\mathcal K}}

\newcommand{\mcV}{{\mathcal V}}

\def \Nat{\mathbb{N}}

\def \N {\Nat}
\def \Sphere{\mathbb{S}}

\title{\textbf{\sc{Gluing construction of initial data with Kerr-de Sitter
ends}}}
\author{Julien Cortier\footnote{Max-Planck-Institut f\"ur Gravitationsphysik
(Albert-Einstein-Institut), Am M\"uhlenberg 1, 14476 Golm, Germany, \texttt{jcortier@aei.mpg.de}.}}
\date{}
\begin{document}

\maketitle

\begin{abstract}
We construct initial data sets which satisfy the vacuum constraint
equations of General Relativity with positive cosmologigal constant.
More presilely, we deform initial data with ends asymptotic to Schwarzschild-de Sitter to obtain
non-trivial initial data with exactly Kerr-de Sitter ends. The method is inspired from
Corvino's gluing method. We obtain here a extension of a previous result for the
time-symmetric case by Chru\'sciel and Pollack in~\cite{CP08}.
\end{abstract}

\tableofcontents

\section{Introduction}
Let $(\mcM,\bar g)$ be a space-time, solution of the vacuum Einstein
equations with cosmological constant $\Lambda$. Let $M$ be a spacelike hypersurface of
this space-time. Then the induced Riemannian metric $g$ and second
fundamental form $k$ on $M$ must satisfy the \emph{constraint equations}:
\begin{equation}
\label{contraintes} \Phi (g,k) := \left(\begin{array}{cc}R(g) -
2\Lambda - |k|^2_{g} + (\tr _g k)^2  \\ 2\left(\Div_g k - d(\tr _g
k)\right)
\end{array}\right) = 0 \;.
\end{equation}
This set of equations is obtained by the Gauss-Codazzi-Mainardi
formulas (see for example~\cite{BI04}), and one immediately notices
that these are intrinsic, i.e. depend only on the induced data on
$M$.

On the other hand, a major result of Choquet-Bruhat in~\cite{FB52}
and Choquet-Bruhat-Geroch in~\cite{CBG69} asserts that, given a set
$(M,g,k)$ of initial data satisfying the constraint equations
(\ref{contraintes}), where $(M,g)$ is a Riemannian manifold and $k$
a symmetric 2-tensor on $M$, one can then obtain the existence of a
space-time $(\mcM,\bar g)$, solution of the vacuum Einstein equations
with cosmological constant $\Lambda$, and an isometric embedding of
$(M,g,k)$ as a spacelike hypersurface of $\mcM$ such that $g$ and
$k$ are respectively the first and the second fundamental form of
the hypersurface in $(\mcM,\bar g)$.

A way of understanding the Einstein equations of general relativity
is therefore to understand the constraint equations themselves.
There is already a large literature that explores the set of
solutions of the constraint equations~(\ref{contraintes}). In
particular, two methods have been fruitful, the \emph{conformal
method} on the one hand, described and used in articles such as \cite{AC94,theseromain,BI04,Hum10}, 
and the \emph{gluing method} on the other hand,
initiated in 2000 by Corvino in~\cite{Cor00}, and which is at the
center of the present article.

Corvino was able to produce in~\cite{Cor00} families of initial data
which are \emph{time-symmetric}, i.e. of the form $(g,0)$, with $g$
asymptotically flat, scalar-flat, with exactly Schwarzchildean ends,
without being forced to be globally Schwarzschild initial data.
Moreover, the decay at infinity required for the asymptotically flat metric to be glued with a Schwarzchildean metric
is the same that the one which ensures that the ADM mass is a well-defined geometric invariant, see also~\cite{Bar86}.
Such a result is therefore of particular interest in problems such as the proof of the Positive Mass Theorem, 
since it makes sufficient to prove the positivity for metrics which, beyond the assumptions of the Theorem, are also
Schwarzschildean at infinity.

In 2006, the gluing method has been used and improved by Corvino and Schoen
in~\cite{CS06} and by Chru\'sciel and Delay in~\cite{CD03} for the
full constraint operator $\Phi$ defined in equation
(\ref{contraintes}), again for asymptotically flat initial data.

Some more recent work has also been done, in the case of a non-zero
cosmological constant $\Lambda$. See~\cite{CD09} for the case
$\Lambda < 0$ (asymptotically hyperbolic, time-symmetric initial
data) and~\cite{CP08} for the case $\Lambda > 0$, with a result
generalized in~\cite{CPP09}. All these results are however
restricted to time-symmetric initial data.

\medskip

In this work, we aim at extending some of the previous results in
the case of a positive cosmological constant to the general initial
data $(g,k)$, thus involving the full constraint operator $\Phi$.
The understanding of space-times with positive cosmological constant
is of special interest in astrophysics since observations such
as~\cite{Per99} suggest that our universe enters in this case.

After a study of some properties of the constraint operators, and
more particularly of its linearization evaluated at the
Schwarzschild-de Sitter reference initial data in
section~\ref{sec_constraint}, we go back in section~\ref{sec_kds} to
the properties of Kerr-de Sitter space-times already pointed out
in~\cite{GH77}, in particular their induced initial data on a
spacelike hypersurface. The section~\ref{sec_deform} aims at
constructing a solution to an auxiliar perturbation problem projected on a
suitable linear space that makes the constraint operator becomes a local diffeomorphism.
This linear space is chosen to be transverse to the cokernel $\mcK_0$ of the linearized constraint operator, also known as the space
of Killing Initial Data (KID). 
Eventually, the section~\ref{sec_solve} focuses on the projection of the problem on
$\mcK_0$ by the means of the global charges associated to the
various initial data involved, therefore completing the proof of
Theorem~\ref{thmrecoll} stated in section~\ref{sec_deform}. As for
gluing problems already treated in~\cite{Cor00,CD03,CS06,CP08}, the
key point consists in using a suitable reference family of initial
data (here the Kerr-de Sitter family) parametrized by as many
independent variables as there are dimensions in the space $\mcK_0$.

For the static case of~\cite{CP08}, the object of study was the
cokernel of the linearized scalar curvature evaluated at a
Schwarzschild-de Sitter metric, which was proved there to be
one-dimensional, and the chosen family of reference metric was
merely the family of Schwarzschild-de Sitter metrics, parametrized by the
mass number $m$. Here, we will see in section~\ref{contraintes} that
since the cokernel $\mcK_0$ has dimension 4, so that one cannot
expect to be able to deform general asymptotically Schwarzschild-de Sitter
initial data to initial data with an exactly Schwarzschild-de Sitter end.
Instead, one can only expect to deform it to initial data which have
an end being exactly an element of a reference family of initial
data with 4 parameters; here the family of Kerr-de Sitter initial data plays
this role, parametrized by the mass $m$ and a vector $\vec a \in \mathbb R^3$ which is related to
the angular momentum.

\paragraph{Acknowledgments} The author wishes to thank Piotr T. Chru\'sciel and Marc Herzlich --who supervised his PhD thesis-- 
and Erwann Delay for useful comments, the Institut de Math\'ematiques et de Mod\'elisation de Montpellier, Universit\'e Montpellier 2, 
and the Albert-Einstein-Institut in Potsdam for financial support.

\section{Some facts about the constraint operator}
\label{sec_constraint} Let us write first the constraint
equations~(\ref{contraintes}) in a slightly different form:
\begin{equation}
\label{contraintes2} \Phi (g,k) := \left(\begin{array}{cc}R(g) -
2\Lambda - |k|^2_{g} + (\tr _g k)^2  \\ 2\nabla_i \left( k^{ij} -
(\tr _g k)g^{ij}\right)
\end{array}\right) = 0 \;.
\end{equation}
$g$ is here a Riemannian metric on a smooth $n$-dimensional manifold $M$,
while $k$ is a symmetric $(0,2)$-tensor on $M$. We are interested in
the sequel in the linearization $D\Phi_{(g_0,k_0)}$ of the operator
$\Phi$ evaluated at a couple $(g_0,k_0)$ of initial data. In
particular, the main object of study is its formal $L^2(g_0)$ adjoint
$D\Phi^*_{(g_0,k_0)}$, which writes:
\be
D\Phi_{(g_0,k_0)}^*(f,Z) = \left(\begin{array}{cc} L_{g_0}^* f - \frac12 \nabla_p(k_0^{pq}Z_q)g_0 + \frac12 \nabla_p (Z^p k_0) \\
     - \frac12 \mathcal L _Z g_0 + (\Div _{g_0} Z)g_0 - 2fk_0 + 2f (\tr _{g_0} k_0)g_0
\end{array}\right)\;,
\ee
where $L_{g_0}^* f$ is the adjoint of the linearized scalar
curvature operator $R$ evaluated at $g_0$ and computed against $f$, and whose expression is:
$$
L_{g_0}^*f = -(\Delta_{g_0} f) g_0 + \mathrm{Hess}_{g_0} f - f
\mathrm{Ric}_{g_0} \;,
$$
where the arguments $f$ and $Z$ are respectively a function and a
vector field on $M$.
The elements $(f,Z)$ of the kernel of $D\Phi^*_{(g_0,k_0)}$ are usually referred to as the \emph{Killing initial data},
or \emph{KIDs} of $(g_0,k_0)$, see~\cite{BC97} (where this terminology was first introduced), and~\cite{Mae04} for an improvement
of this formalism. 
We are now interested in the case where the initial data $(g_0,k_0)$ are the data induced 
by a Schwarzschild-de Sitter (SdS) space-time.
The space-time metric in local coordinates takes the form
\be
\bar g = -V(r)dt^2 + \frac{dr^2}{V(r)} + r^2 h_0\;,
\ee
where $V(r)= f(r)^2 = 1-\frac{2m}{r} - r^2$, and where $h_0$ is the canonical round metric on the unit sphere $\Sphere^2$.
This expression is valid on a range $(r_b,r_c)$ of $r$ where $V(r)$ is positive, where $r_b$ and $r_c$ are the two positive roots of $V(r)$.

These space-times are static, so that the initial data induced on the hypersurface $\{t=0 \}$ write on the form $(g_0,0)$.
This leads to the simplification
\be
D\Phi_{(g_0,0)}^*(f,Z) = \left(\begin{array}{cc} L_{g_0}^* f  \\
     - \frac12 \mathcal L _Z g_0 + (\Div _{g_0} Z)g_0
\end{array}\right)\;.
\ee
To compute the kernel of the operator above, one already knows that the functions $f$ on $M$ such that 
$L_{g_0}^* f =0$ form a one-dimensional space
when the mass parameter $m$ is non-zero, as states the Lemma 3.3 p 647 of~\cite{CP08}. 
On the other hand, one has to look for vector fields $Z$
tangent to $M$ such that $- \frac12 \mathcal L _Z g_0 + (\Div _{g_0} Z)g_0 = 0$. 
But these are exactly the Killing vector fields for the metric $g_0$,
since the equation above, when computing its trace, reduces to $\mathcal L _Z g_0 = 0$. 
Since the Killing fields of $g_0$ form a three-dimensional space,
we eventually find that the space of \emph{KIDs} of $(g_0,0)$ form a space of dimension 4. 
In the sequel, we denote this space $\mcK_0$, and we write it
$$\mathcal K_0 = \mathrm{Vect}\left(V_0,V_1,V_2,V_3\right)\;,$$
with $V_0 = (f,0)$ where $f(r) = \sqrt{V(r)}$, $V_i = x^j \frac{\partial}{\partial x^k} - x^k \frac{\partial}{\partial x^j}$, 
for $(i,j,k)$ being a 3-cycle of $(1,2,3)$.

We therefore aim at finding a 4-parameter family of space-times in
which one could chose an element to perform a gluing construction
with initial data that asymptote to a spacelike slice of a SdS
space-time; a natural candidate for this family is the Kerr-de Sitter
family of space-times.

\section{Kerr-de Sitter space-times and their initial data}
\label{sec_kds}
\subsection{Space-time metric and analytic extension}

\begin{figure}[htbp]
\begin{center}
\includegraphics[width= 0.5\linewidth]{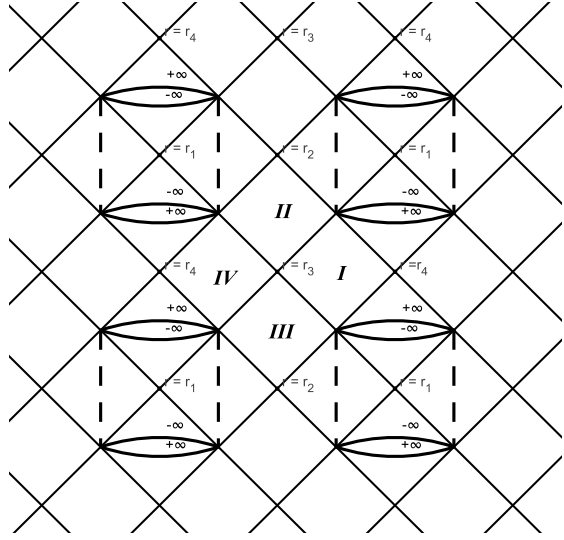}
\caption{Carter-Penrose diagram of the Kerr-de Sitter space-time
showing causal relations between various regions of the analytic extension.
\newline
Vertical dashed lines represent the locus of the singularity $\{r =
0, \theta = \pi/2\}$. } \label{diagkkds}
\end{center}
\end{figure}

Kerr-de Sitter metrics, also labeled KdS in the sequel, form a
family parametrized by numbers $m$ and $a$ of solutions of the
vacuum Einstein equations with positive cosmological constant. In
the Boyer-Lindqvist local coordinate system, the KdS metrics take the form:
\bea \bar g_{m,a} = \rho^2
\left(\frac{dr^2}{\Delta_r}+\frac{d\theta^2}{\Delta_{\theta}}\right)
+ \sin^2 \theta \frac{\Delta_{\theta}}{\rho^2}\left(
\frac{a dt - (r^2 + a^2)d \varphi}{1 + \Lambda a^2/3}\right)^2 \\
\nonumber{- \frac{\Delta_r}{\rho^2}\left(\frac{dt - a \sin^2 \theta
d \varphi}{1 + \Lambda a^2/3}\right)^2}\;, \eea
with
$$\Delta_r = (r^2 + a^2)(1-\frac{\Lambda r^2}{3}) - 2 m r\;,\
\Delta_{\theta} = 1 + \frac{\Lambda}{3}a^2 \cos^2 \theta\;,\
\rho^2 = r^2 + a^2 \cos^2 \theta\;.$$
In what follows, we choose sufficiently small values of the
parameters $m>0$ and $a$ such that $\Delta_r$ admits four distinct
real roots, which are noted
$$r_1 < r_2 < r_3 < r_4\;.$$
The above coordinates $\theta$ and $\varphi$ are usual spherical
coordinates on the unit 2-sphere, while $r$ takes values in the
interval $(r_3,r_4)$ on which $\Delta_r$ is positive. In this range
of values for $r$, the metric $\bar g_{m,a}$ is Lorentzian, analytic
and satisfies the vacuum Einstein equations in dimension 3+1 with
positive cosmological constant $\Lambda$. The above expression in
coordinates $(t,r,\theta,\varphi)$ gives rise to an apparent
singularity at both $r=r_3$ and $r=r_4$. It is however possible to
construct an analytic extension of the metric across the horizons
$\{r=r_3\}$ (black hole event horizon) and $\{r=r_4\}$ (cosmological
event horizon). The construction, already mentioned in~\cite{GH77},
is the following: we define new coordinates $u$ and $\hat{\varphi}$
by:
$$du = dt + \frac{\lambda}{r-r_3}dr\ ,\ d\hat{ \varphi} = d\varphi - \mu dt\;,
$$
where $\lambda$ and $\mu$ are constants to be determined. Then we
study the coefficients of the metric $\bar g_{m,a}$ in coordinates
$(u,r,\theta,\hat\varphi)$. In particular, the coefficient $\bar
g_{rr}$ becomes:
\beal{bargrr}
\bar g_{rr} = -\frac{\Delta_r}{A^2 \rho ^2}\frac{\lambda ^2}{(r-r_3)^2}(1 - a \mu \sin^2 \theta )^2 + \frac{\rho^2}{\Delta_r} \\
\nonumber{+ \sin^2 \theta \frac{\Delta_{\theta}}{A^2
\rho^2}\frac{\lambda^2}{(r-r_3)^2}(a - \mu (r^2 + a^2))^2}\;, \eea
where we have introduced the constant $A := 1 +
\frac{\Lambda}{3}a^2$. The first line of (~\ref{bargrr}) seems to
have a first-order pole at $r=r_3$, whereas the second line seems to
have a second-order pole there. But when one chooses
\bel{valeurmu} \mu = \frac{a}{r_3^2 + a^2}\;, \ee
the singularity at $r=r_3$ in (~\ref{bargrr}) disappears from the
second line of this expression, whereas the cancelation of the
singular part of the first line of the equation evaluated at $r=r_3$
requires the choice of $\lambda$ such that
\bel{valeurlambda} \lambda^2 = \frac{9}{\Lambda^2}\frac{A^2 (r_3^2 +
a^2)^2}{(r_4 - r_3)^2 (r_3 - r_2)^2 (r_3 - r_1)^2}\;. \ee
One can check that these values also ensure that the other metric
coefficients of $\bar g = \bar g_{m,a}$ are also bounded in the
neighborhood of $\{r=r_3\}$ in the
$(u,r,\theta,\hat\varphi)$-coordinate system.We can therefore
analytically extend $\bar g$ to a space-time metric, solution of the
vacuum Einstein equations with cosmological constant $\Lambda$ on
the whole region $\{r_2 < r < r_4 \}$. This region is represented as
the union of the two regions labeled \textbf{\emph{I}} and
\textbf{\emph{II}} of the diagram of the Figure~\ref{diagkkds}.
One can similarly introduce a new coordinate $v$ instead:
$$dv = dt - \frac{\lambda}{r-r_3}dr\;,
$$
where $\lambda > 0$ is again given by the formula~(\ref{valeurmu});
the metric coefficients obtained in the coordinate system $(v,r,\theta,\hat\varphi)$
are again bounded at $r=r_3$, and $\bar g$ can be analytically extended through the
horizon $\{r=r_3\}$ as done previously (regions \textbf{\emph{I}} and
\textbf{\emph{III}} of the diagram of the Figure~\ref{diagkkds}).

It is now tempting to write the metric in coordinates $(u,v,\theta,\hat\varphi)$ in
order to overlap the two previous extensions, but the metric $\bar g$ becomes
degenerate at $r=r_3$ in this coordinate system. A classical method~(\cite{Kru60,Car68,HE73}) to avoid this is
to set \emph{exponential coordinates} defined as:
$$\hat u := e^{cu}\ ,\ \hat v := - e^{-cv}\;,$$
where $c$ is a constant to be suitably chosen later. The smoothness and non-degeneracy of the metric $\bar g$ at
$r=r_3$ imposes the choice
\bel{valeurc}
c = \frac{1}{2\lambda}\;.
\ee
Since $r$ can therefore be expressed as a smooth function of $\hat u$ and $\hat v$ by the identity
$$\hat u \hat v = -(r-r_3)\;,$$
we obtain that $\bar g$, initially defined in the region $\{\hat u > 0\ ,\ \hat v < 0\ ,\ r_3 < r < r_4\}$,
analytically extends to a smooth Lorentzian metric, solution of
the vacuum Einstein equations with cosmological constant $\Lambda$
to the region $\{\hat u \in \mathbb R\ ,\ \hat v \in \mathbb R\ ,\ r_2 < r < r_4\}$, through the horizon
$\{r=r_3\}$ which presents there a bifurcation structure as it also reads $\{\hat u \hat v = 0\}$.
On the diagram of Figure~\ref{diagkkds}, this extended space-time corresponds to the union of the four regions
labeled from \textbf{\emph{I}} to \textbf{\emph{IV}}.

\medskip

One can repeat this process to construct an analytic extension of $\bar g$ across the \emph{cosmological}
horizon (see~\cite{GH77}) $\{r=r_4\}$, by choosing again suitable values of the corresponding constants
$$\tilde \mu = \frac{a}{r_4^2 + a^2}\;,
$$
and
$$\tilde \lambda^2 = \frac{9}{\Lambda^2}\frac{A^2 (r_4^2 + a^2)^2}{(r_4 - r_3)^2 (r_4 - r_2)^2 (r_4 - r_1)^2}\;.
$$
One can further construct analytic extensions across the horizons $\{r=r_2\}$ (``inner'' black hole event horizon)
and $\{r=r_1\}$ (``inner'' cosmological horizon), and, by induction, we obtain a maximal analytic extension
$(\widehat{\mathcal M}_{m,a},\hat{g}_{m,a})$ of the initial space-time $(\mcM_{m,a},\bar g_{m,a})$.
The Figure~\ref{diagkkds} shows the extended space-time and the various horizons.
\begin{engrk}
 At this stage it is important to note that the above Kerr-de Sitter metric with parameters $m$ and $a$ has an angular momentum vector
$\vec a = a \partial_z$, hence directed along $\partial_z$, where $z$ is one of the cartesian coordinates $x,y,z$ with respect to which
the spherical coordinates $r,\theta,\varphi$ are defined.
\end{engrk}

\subsection{Induced initial data}
From the above Carter-Penrose diagram, one expects that the induced initial data on a slice $\Sigma := \{t=0\}$
(or more precisely on its analytic extension) are periodic; we will see later that we can indeed define in a natural way 
a coordinate $y$ on $\R$, such that the coordinate $r$ restricted to $\Sigma$ is a smooth periodic function of $y$,
taking values in the set $[r_3,r_4]$.

From now, we will only consider the case $\Lambda =3$; the general case $\Lambda > 0$ can be deduced from that by a scaling.
We will denote $(\mathring g_{m,a},\mathring k_{m,a})$ the initial data on $\Sigma$. 
In the $(r,\theta,\varphi)$-coordinate system, the induced metric $\mathring g_{m,a}$ reads
$$\mathring g_{m,a} = \rho^2
\left(\frac{dr^2}{\Delta_r}+\frac{d\theta^2}{\Delta_{\theta}}\right)
+ \frac{\sin^2 \theta}{\rho^2(1+a^2)^2}\left(\Delta_{\theta}(r^2 + a^2)^2 - a^2 \sin^2\theta \Delta_r\right)
d \varphi^2\ ,
$$
and one can check that the coefficients of $\mathring k_{m,a}$ are zero 
except $k_{r\varphi}=k_{\varphi r}$ and $k_{\theta \varphi}=k_{\varphi \theta}$.
In particular, we write here the expression of $k_{r\varphi}$ which will be of use of later:
\bel{krphi}k_{r\varphi} = \frac{a m \sin^3\theta \sqrt{\Delta_{\theta}}\left(a^2 (a^2 - r^2) \cos^2\theta - r^2 a^2 - 3r^4\right)}
{\rho^3 (1+a^2) \sqrt{\Delta_r}\sqrt{\Delta_{\theta}(r^2 + a^2)^2 - \Delta_r a^2 \sin^2 \theta}}\;,
\ee
whereas the expression of the $(\theta,\varphi)$-component is
$$k_{\theta \varphi} = \frac{4 a^3 m \cos\theta \sin^5\theta \sqrt{\Delta_r}\sqrt{(r^2 + a^2)(1+a^2) + 2 a^2 m r \sin^2\theta}}
{\rho (1+a^2)^3\sqrt{\Delta_{\theta}}}\ . 
$$

We now aim at making explicit the periodicity property of these initial data. In the Schwarzschild-de Sitter case,
 Chru\'sciel and Pollack in~\cite{CP08} showed that the induced metric on the slice $\{t=0\}$,
$$b _m = \frac{dr^2}{V(r)} + r^2 g_{\mathbb S^2}$$
is exactly a Delaunay metric of the form
$$\mathring g _{\varepsilon} = u^4 _{\varepsilon}(y)\left(dy^2 + g_{\mathbb S^2}\right)\;,$$
where $y$ is defined on $\mathbb R$, and where $u_{\varepsilon}$ is a positive solution of
$$u'' - \frac14 u + \frac34 u^5 = 0\ ,\ u(0)=\varepsilon = \min u\;,$$
with $\varepsilon$ belonging to the interval $(0,\bar {\varepsilon})$, with $\bar {\varepsilon} = \left(\frac13\right)^{1/4}$.
Note that the number $\varepsilon$ corresponds to the \emph{neck size}, namely the minimal distance from the rotation axis of the 
Delaunay manifold.
The positive solutions of this equation are studied by Schoen in~\cite{Smontecatini} and by Mazzeo, Pollack and Uhlenbeck in~\cite{MPU96}.
They are shown to be periodic, with a period depending on $\varepsilon$.

The way to switch coordinates between $r$ and $y$ is there the following:
\bel{rtoy}
\frac{dy}{dr} = \frac{1}{\sqrt{r^2 - 2 m r - r^4}} = \frac{1 }{r \sqrt{V(r)}}\;,
\ee
so that $y$ is well defined as a function (elliptic integral) of $r$ on every interval on which $V$ is positive,
in particular on $(r_b,r_c)$, where $r_b$ and $r_c$ are the two positive roots of the polynomial $rV(r) = r - 2m - r^3$.
On the other hand, one has $r=u_{\varepsilon}^2$, so that $r$ is actually a function of $y$ defined on $\R$,
which is moreover smooth, periodic, even, so that the period is obtained by the formula:
$$T(\varepsilon) = 2 \int_{r=r_b}^{r_c}\frac{dr}{r \sqrt{V(r)}}\;.$$
Note that the parameters $m$ and $\varepsilon$ are related by the identity $\varepsilon^2 = r_b (m)$, so that the period
$T(\varepsilon)$ varies with the mass parameter $m$. 

\medskip

We now use a similar approach to find coordinates in which the periodicity of the Kerr-de Sitter initial data is made obvious.
Recall that the Riemannian metric $\mathring g_{m,a}$ induced by $\bar g_{m,a}$ on $\Sigma$ writes:
$$
\mathring g _{m,a} = \rho^2 \left(\frac{dr^2}{\Delta_r}+\frac{d\theta^2}{\Delta_{\theta}}\right)
+ \frac{\sin^2 \theta}{\rho^2 (1 + a^2)^2}\left[\Delta_{\theta}(r^2 + a^2)^2 - \Delta_r a^2 \sin^2 \theta\right]d\varphi^2\;,$$
so that we set
\bel{rtoyKKdS}
\frac{dy}{dr} = \frac{1}{\sqrt{\Delta_r}}\ ,\ \rho^2 =: u^4_a (y,\theta)\;.
\ee
This change of variables allows us to define $y$ as a smooth function of $r$ on the interval $(r_3,r_4)$ where $\Delta_r$ is positive.
By morevover adding the choice of the origin $r(y=0) = r_3$, we also obtain $r$ as an elliptic function of $y$, which therefore extends
by periodicity and parity to a smooth function defined on $\R$.

Hence, the metric $\mathring g _{m,a}$ can be expressed in coordinates $(y,\theta,\varphi)$ as
$$\mathring g_{m,a} = u^4_a(y,\theta) \left(dy^2 + \frac{d\theta^2}{\Delta_{\theta}}\right) +
\frac{\sin^2 \theta}{\rho^2 (1 + a^2)^2}\left[\Delta_{\theta}(r^2 + a^2)^2 - \Delta_r a^2 \sin^2 \theta\right]d\varphi^2\;,$$
namely in a periodic metric with respect to $y$, with period $T(m,a)$ given by the integral
\bel{T(m,a)}
T(m,a) = 2 \int_{r_3}^{r_4} \frac{dr}{\sqrt{\Delta_r}}\;.
\ee
One can likewise write the second fundamental form $\mathring k_{m,a}$ of $\Sigma$ in these coordinates to make it
periodic with the same period with respect to $y$.
Note that the period $T(m,a)$ does vary with $m$ and $a$, in a continuous way from the formula~(\ref{T(m,a)}) above.

\section{Construction of deformed initial data}
\label{sec_deform}
This section, together with the next one, aims at proving the following Theorem which is the main result in this article:
\begin{Theorem}\label{thmrecoll}
 Let $(g,k) \in \Gamma(\mcM_{et} \times S^2 T^*M)$ be a pair of initial data on a smooth non-compact 
3-manifold $M$ which solves the constraint equations
$\Phi(g,k)=0$. Assume that $M$ contains a region $M_{\infty} = [R_0,\infty) \times \Sphere^2$ on which the initial data $(g,k)$ asymptote
to the induced initial data $(g_0,k_0) = (b_{m_0},0)$ of a Schwarzschild-de Sitter space-time of parameter $m_0$. 
More precisely, we require that the $g_0$-norms of $g-g_0$ and of $k-k_0$, together with their derivatives up to some finite order,
tend to 0 as $y \rightarrow +\infty$.
Then there exists a number $m >0$, a vector $\vec a$ in $\R^3$ and a pair of initial data $(\tilde g, \tilde k)$ solution of
$\Phi(\tilde g, \tilde k) = 0$, which coincides with $(g,k)$ on a compact subset of $M$ and which also coincides with the induced 
initial data
of a Kerr-de Sitter space-time $(\mathring g _{m,\vec a},\mathring k _{m,\vec a})$ on the complement of a compact set of $M_{\infty}$.
\end{Theorem}
\noindent Here, $\mcM_{et}$ is the space of Riemannian metrics on $M$.

\begin{figure}[htbp]
\begin{center}
\includegraphics[width= 0.6\linewidth]{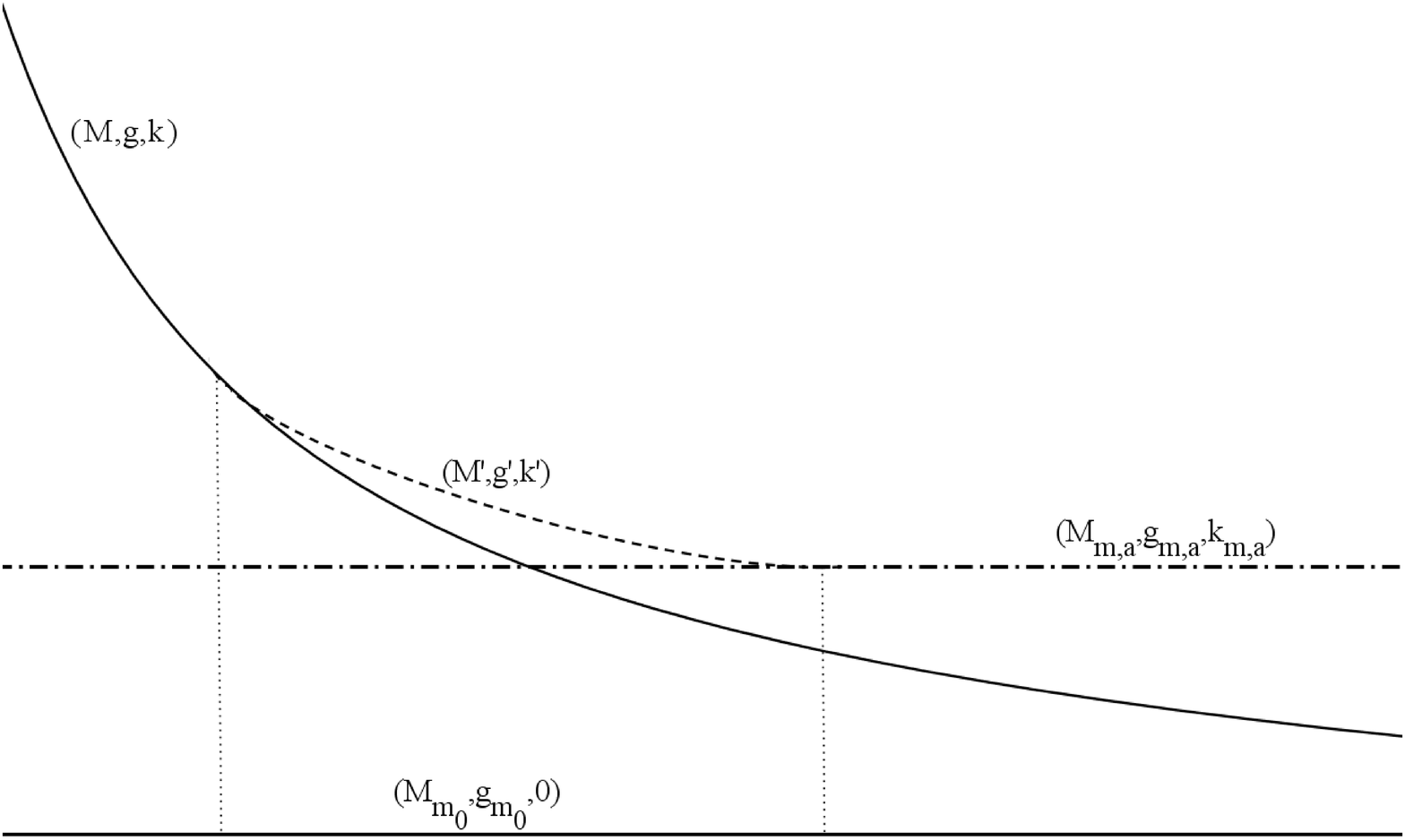}
\caption{Gluing principle.
\newline
The initial data $(M,g,k)$ are represented by the curve which
asymptotes to the horizontal line referring to the initial data
$(M_{m_0},\mathring g_{m_0},0)$. We glue it with initial data of
$(M_{m,a},\mathring g_{m,a},\mathring k_{m,a})$ (represented by the dashed horizontal line)
to obtain $(M',g',k')$. The gluing zone is located between the two vertical dashed segments.}
\label{gluing}
\end{center}
\end{figure}

\begin{figure}[htbp]
\begin{center}
\includegraphics[width= 0.9\linewidth]{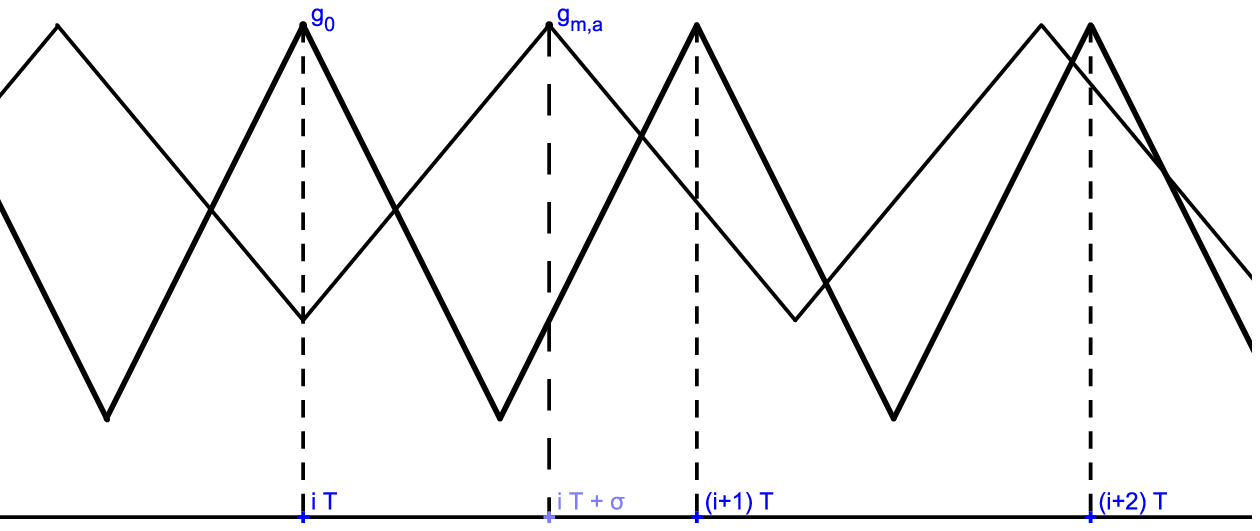}
\caption{Illustration of the definition of $\sigma = \sigma (i,T,T')$.
\newline
The fat saw-toothed line represents the reference initial data $(g_0,0)$ which are
$T$-periodic with respect to the variable $y$, whereas the thin saw-toothed line represents
the $T'$-periodic initial data $(\mathring g_{m,\vec a},\mathring k_{m,\vec a})$.}
\label{defsigma}
\end{center}
\end{figure}

Let $(g,k)$ satisfy the assumptions of the above Theorem. Without
loss of generality, we will now assume that $R_0 = 0$. We denote
$T=T(m_0)$ the period of $g_0$ with respect to the $y$-coordinate
described in the previous section. For every admissible $(m,\vec a)$
in $\R^4$, every $i$ in $\N$ and $\sigma$ in $[0,T)$, we start by
interpolating, by the means of a cut-off function, between $(g,k)$
and $(\mathring g _{m,\vec a},\mathring k _{m,\vec a})$ in the
region $\Omega_{i,\sigma}$, where
$$\Omega_{i,\sigma} = \left(iT + \sigma + \delta , (i+1)T + \sigma - \delta\right) \times \mathbb S^2\;,$$
where $\delta$ is a constant in $(0,T/3)$. $\Omega_{i,\sigma}$ is therefore a product of an interval of length $T-2\delta$ with $\Sphere^2$.
Note in particular that the function $f(r) = \sqrt{V(r)}$ remains positive on the compact set $\overline{\Omega_{i,\sigma}}$.
The interpolation reads:
\bea
g_{m,\vec{a},i,\sigma} := (1- \chi_{i,\sigma})g + \chi_{i,\sigma} \mathring g_{m,\vec{a}}\;, \\
\nonumber{k_{m,\vec{a},i,\sigma} := (1- \chi_{i,\sigma})k + \chi_{i,\sigma} \mathring k_{m,\vec{a}}\;,}
\eea
and is illustrated in Figure~\ref{gluing}. The functions $\chi_{i,\sigma}$ are cut-offs, compactly supported in
$(iT + \sigma + \delta, (i+1)T + \sigma - \delta)$. More precisely, all these cut-off functions are defined from a single one,
whose support is translated: we namely take $\chi_{i,\sigma}(y)= \chi(y - iT - \sigma)$, where $\chi$ is a smooth function
having values in $[0,1]$ which is zero on $[0,T/3]$ and equal to one on $[2T/3,T]$.
From the asymptotic assumptions on the initial data $(g,k)$, we will be able to make the image of the interpolated initial data
$g' = g_{m,\vec{a},i,\sigma}$ and $k'= k_{m,\vec{a},i,\sigma}$ by the constraint operator $\Phi$ as small as needed, provided that
one choose a parameter $i$ big enough, and suitable values of $m,\vec a$ and $\sigma$. But for further analysis,
we will need a fixed gluing zone which does not depend on $i$.
We can avoid this difficulty by defining the following quantities, obtained from a translation of $(g,k)$:
$$g_{i,\sigma}(y,\theta,\varphi) := g(y + iT + \sigma, \theta, \varphi)\ ;\ k_{i,\sigma}(y,\theta,\varphi) := k(y + iT + \sigma,
\theta, \varphi)\;.$$
We similarly define $\mathring g_{m,\vec{a},i,\sigma}$ and $\mathring k_{m,\vec{a},i,\sigma}$
from $\mathring g_{m,\vec{a}}$ and $\mathring k_{m,\vec{a}}$ respectively.
We also make the choice
$$\sigma := \sigma(i,T,T') := iT' \mod T \ \in [0,T) \;,$$
so that the metrics $\mathring g_{m,\vec{a},i,\sigma}$ and $g_0$, periodic with respective periods $T'$ and $T$, have the same phase
at $y=0$, see Figure~\ref{defsigma}.
With this condition on $\sigma$, the initial data $(\mathring g_{m,\vec a,i,\sigma},\mathring k_{m,\vec a,i,\sigma})$
do not depend on $i$ and equal $(\mathring g_{m,\vec a},\mathring k_{m,\vec a})$ on $\Omega$.
We then replace  $g_{m,\vec{a},i,\sigma}(y)$ by $g_{m,\vec{a},i,\sigma}(y + iT + \sigma)$
and $k_{m,\vec{a},i,\sigma}(y)$ by $k_{m,\vec{a},i,\sigma}(y + iT + \sigma)$.
The above value imposed to $\sigma$ makes it depend on $i,m,\vec a$. We therefore adopt the following definitions:
\beal{interpolation}
g' := g_{m,\vec{a},i} := (1- \chi)g_{i,\sigma} + \chi \mathring g_{m,\vec{a}}\;, \\
\nonumber{k' := k_{m,\vec{a},i} := (1- \chi)k_{i,\sigma} + \chi \mathring k_{m,\vec{a}}\;.}
\eea
Hence, the gluing zone, as for the cut-off function, is compactly supported in $\Omega$. Moreover, $\bar \Omega$
is a smooth manifold with boundary $\partial \Omega = (\{\delta\}\times \mathbb S^2) \cup (\{T- \delta\}\times \mathbb S^2)$.
We take a defining function $x$ of the boundary $\partial \Omega$, for example
$$x(y) := \frac{T- 2\delta}{2\pi}\sin \left(\frac{2\pi}{T-2\delta}(y - \delta)\right)\;.$$
Note at this point that the interpolation $(g',k')$ satisfies the constraint equation outside $\Omega$, but has no reason
to do so in $\Omega$. However, from a heuristic point of view, $(g',k')$ will be a ``good'' approximate solution 
of the constraint equations for large $i$. 
We wish now to find a small perturbation $(\delta g,\delta k)$ such that $\Phi(g' + \delta g,k' + \delta k)=0$.
However, the fact that the cokernel $\mcK_0$ of $D\Phi_{(g_0,0)}$ is not trivial motivates us to first look at the projection 
of the problem on a subspace transerve to $\mcK_0$. 
We will then use the following classical result, which is a consequence of the Picard's fixed point Theorem:

\begin{Lemma}\label{ptfixe} Let $\Psi : E \rightarrow F$ a smooth map between two Banach spaces $E$ and $F$. Let $Q : E \rightarrow F$
defined by $Q(x) = \Psi(x) - \Psi(0) - D\psi(0)\cdot x$ . Assume that there exists positive constants $q$, $r_0$, $c$ such that:
\begin{enumerate}
\item $\forall x,y \in B_E(0,r_0)\ ,\ \|Q(y) - Q(x)\| \leq q \|y-x\| (\|x\| + \|y\|)\;,$
\item $D\Psi(0)$ is invertible, and the inverse is bounded by $c$\;.
\end{enumerate}
Then, for all $r > 0$ such that $r < \min \left(\frac{1}{2qc}, r_0\right)$, if $\|\Psi(0)\| \leq \frac{r}{2c}$, there exists a unique
$x \in B_E(0,r)$ solution of $\Psi(x) = 0$.
\end{Lemma}
%


We wish to apply this Lemma in the following way:

\noindent Recall that for all $i \in \N$, $(m,\vec a) \in \R ^4$, $(g',k')$ is the interpolation of initial data as described above. 
We define
$\Psi_{i,m,\vec a} : E(g') \longrightarrow F(g')$, by the formula
$$\Psi_{i,m,\vec a}(N,Y) := \Pi_{\mathcal K _0 ^{\bot g'}}\ e^{-2s/x} \Phi\left((g',k') + (\delta g,\delta k)\right)\;,$$
where
$$(\delta g,\delta k) = \varXi e^{2s/x} \varXi P^*(N,Y)\;,$$
$$E(g') = \mathcal K _0 ^{\bot g'} \cap \left(\mathring H ^{k+4}_{x^2,e^{s/x}}(g') \times \mathring H ^{k+3}_{x^2,e^{s/x}}(g')\right)$$
and
$$F(g') = \mathcal K _0 ^{\bot g'} \cap \left(\mathring H ^{k}_{x^2,e^{s/x}}(g') \times \mathring H ^{k+1}_{x^2,e^{s/x}}(g')\right)\;.$$
The spaces $\mathring H^l_{x^2,e^{s/x}}(g')$ are the weighted Sobolev spaces (with respect to the defining function $x$) 
which are defined in~\cite{CD03}.
In what follows, to avoid to many parameters, we take $s=1/2$. We assume that the parameter 
(not to be confused with the second fundamental form) $k$ is greater than $3/2$, following~\cite{CD03}.
More precisely, $k \geq 0$ suffices for having a linear isomorphism $D\Psi_{i,m,\vec a}$, whereas $k > 3/2$ is required to
have $\Psi_{i,m,\vec a}$ well-defined.
Here $\varXi$ is the map $(u,v) \mapsto (x^4 u, x^2 v)$ (the one that is denoted by $\Phi$ in~\cite{CD03}) and $P$
is $D\Phi(g',k')$, $P^*$ its formal adjoint. $E(g')$ and $F(g')$ are Banach spaces.
Hence, the maps $\Psi_{i,m,\vec a}$ are well-defined, differentiable, and, following the
notations of~\cite{CD03}, 
$$D\Psi_{i,m,\vec a}(0,0) = \Pi_{\mathcal K _0 ^{\bot g'}}\ e^{-2s/x} P \varXi e^{2s/x} \varXi P^* =: 
\Pi_{\mathcal K _0 ^{\bot g'}}L_{x^2,e^{s/x}} \;,$$
hence by the  Theorem 3.6 of~\cite{CD03}, $D\Psi_{i,m,\vec a}(0,0)$ is an isomorphism between $E(g')$ and $F(g')$, such that the inverse
\bel{DPsiinverse}\| D\Psi_{i,m,\vec a}(0,0)^{-1} \|_{F(g'),E(g')} \leq C \ee
is bounded independently of $i,m,\vec a$ for $i \geq i_0$, $(m,\vec a) \in B((m_0,\vec 0),\delta_0)$, for some $i_0 \in \N$ and
$\delta_0 > 0$.
On the other hand, for all $i,m,\vec a$, $\Psi_{i,m,\vec a}$ is smooth, in particular is in $W^{2,\infty}_{loc}(g_0)$, thus
there exists $q_{i,m,\vec a} > 0$ and $r_{0,i,m,\vec a} > 0$ such that the condition $1$ of the Lemma~\ref{ptfixe} is satisfied:
$$\|Q_{i,m,\vec a}(y) - Q_{i,m,\vec a}(x)\| \leq q_{i,m,\vec a} \|y-x \|\big(\|x\| + \|y\|\big)\;,
\forall x,y \in B_{E(g')}(0,r_{0,i,m,\vec a})\;, $$
where
$$Q_{i,m,\vec a} = \Psi_{i,m,\vec a} - \Psi_{i,m,\vec a}(0,0) - D\Psi_{i,m,\vec a}(0,0)\;.$$
However, we wish to obtain such an inequality, but with constants
$q$ and $r_0$ which do not depend anymore on $i$ and on $(m,\vec a)$
for $i \geq i_0$ and $(m,\vec a) \in B((m_0,\vec 0),\delta_0)$,
provided that we have increased $i_0$ and decreased $\delta_0$ if
necessary.

\begin{prop}
There exists positive constants $q$ and $r_0$, and $i_0 \in \mathbb N$, $\delta_0 >0$ such that
the inequality 
$$\|Q_{i,m,\vec a}(y) - Q_{i,m,\vec a}(x)\|_{F(g')} \leq q \|y-x \|_{E(g')}\big(\|x\|_{E(g')} + \|y\|_{E(g')}\big)$$
holds, for all $x,y$ in $B_{E(g')}(0,r_0)$, for all $i \geq i_0$ and for all $(m,\vec a)$ such that 
$|(m,\vec a)-(m_0,\vec 0)| < \delta_0$.
\end{prop}

\proof

We will show that we have a uniform convergence of the expression
$$\frac{\|Q_{i,m,\vec a}(y) - Q_{i,m,\vec a}(x)\|_{F(g')}}{\|y-x \|_{E(g')}\big(\|x\|_{E(g')} + \|y\|_{E(g')}\big)}$$
towards 
$$\frac{\|Q_{\infty,m_0,\vec 0}(y) - Q_{\infty,m_0,\vec 0}(x)\|_{F(g_0)}}{\|y-x \|_{E(g_0)}\big(\|x\|_{E(g_0)} + \|y\|_{E(g_0)}\big)}$$
as $(i,m,\vec a) \rightarrow (\infty,m_0,\vec 0)$, uniformly with respect to $x$ and $y$ in some ball of radius $r_0$ 
for the $E(g')$ norm, with $x \neq y$.
We start by showing that one has the above uniform convergence when all the norms are taken to be with respect to the metric $g_0$.
Note that $Q_{i,m_0,\vec 0}$ does not depend on $i$, so one can write $Q_{i,m_0,\vec 0}$ or $Q_{\infty,m_0,\vec 0}$ 
to express the same quantity.
From the definition of $Q$ we can write 
$$\frac{\left|\|Q_{i,m,\vec a}(y) - Q_{i,m,\vec a}(x)\|_{g_0} - 
\|Q_{i,m_0,\vec 0}(y) - Q_{i,m_0,\vec 0}(x)\|_{g_0}\right|}{\|y-x \|_{g_0}\big(\|x\|_{g_0} + \|y\|_{g_0}\big)} \leq
I + II\ ,
$$
where 
$$ I = \frac{\|(\Psi_{i,m,\vec a} - \Psi_{i,m_0,\vec 0})(y) - (\Psi_{i,m,\vec a} - \Psi_{i,m_0,\vec 0})(x)
- D(\Psi_{i,m,\vec a} - \Psi_{i,m_0,\vec 0})(x).(y-x)\|_{g_0}}{\|y-x\|_{g_0}^2}\ ,$$
and 
$$ II = \frac{\|\left[D\Psi_{i,m,\vec a}(x) - D\Psi_{i,m,\vec a}(0)  - 
\left(D\Psi_{i,m_0,\vec 0}(x) - D\Psi_{i,m_0,\vec 0}(0)\right)\right].(y-x) \|_{g_0}}
{\|y-x \|_{g_0}\big(\|x\|_{g_0} + \|y\|_{g_0}\big)}\ .$$
If we denote $\Psi = \Psi_{i,m,\vec a} - \Psi_{i,m_0,\vec 0}$, 
the terms above read
$$I =  \frac{\| \Psi(y) - \Psi(x) -D\Psi(x).(y-x)\|_{g_0}}{\|y-x\|_{g_0}^2}\ ,$$
$$II = \frac{\|\left[D\Psi(x) - D\Psi(0)\right].(y-x) \|_{g_0}}
{\|y-x \|_{g_0}\big(\|x\|_{g_0} + \|y\|_{g_0}\big)} \leq \frac{\| D\Psi(x) - D\Psi(0) \|_{g_0}}{\|x\|_{g_0}}\ .$$
From Taylor-Lagrange inequality, we have
$$ I \leq \frac12\sup _{z \in B_{E(g_0)}(0,r_0)}\| D^2 \Psi (z)\|_{g_0}\ ,$$
and the right-hand term does not depend on $x$ and $y$ in $B_{E(g_0)}(0,r_0)$.
On the other hand, we also have $II \leq \sup _{z \in B_{E(g_0)}(0,r_0)}\| D^2 \Psi (z)\|_{g_0}$.
So, in order to conclude this case, we have to prove that $\sup _{z \in B_{E(g_0)}(0,r_0)}\| D^2 \Psi (z)\|_{g_0}$ goes to $0$
as $(i,m,\vec a)$ goes to $(\infty,m_0,\vec 0)$. 

\medskip
Indeed, note that 
$$\Psi_{i,m,\vec a}(z) = \Pi_{\mathcal K_0 ^{\perp g'}}e^{-\frac1x}\Phi\big(G' + F^2 e^{\frac1x}P^* z \big).
$$
We write 
$$\sup _{z \in B_{E(g_0)}(0,r_0)}\| D^2 \Psi (z)\|_{g_0} \leq III + IV\ ,$$
where
$$III = \|\left(\Pi_{\mathcal K_0 ^{\perp g'}} - \Pi_{\mathcal K_0 ^{\perp g_0}}\right)
e^{-\frac1x}\Phi\big(G' + F^2 e^{\frac1x}P^* z \big) \|_{g_0} $$
and
$$IV = \|\Pi_{\mathcal K_0 ^{\perp g_0}}
e^{-\frac1x}\left[\Phi\big(G' + F^2 e^{\frac1x}P^* z \big) - \Phi\big(G_0 + F^2 e^{\frac1x}P_0^* z \big)\right] \|_{g_0}\ . $$
We show that both expressions go to zero uniformly in $z$ in some ball $B_{E(g_0)}(0,r_0)$ as $(i,m,\vec a)$ 
tends to $(\infty,m_0,\vec 0)$.
For the expression $III$, this comes from the fact that we have the convergence
\bel{convproj}\Pi_{\mathcal K_0 ^{\perp g'}}W \longrightarrow \Pi_{\mathcal K_0 ^{\perp g_0}}W\ee
being uniform with respect to $W$ in a neighborhood of $0$ in the appropriate $g_0$-norm as $g'$ 
tends to $g_0$. Indeed, recall that $\mcK_0$ is finite-dimensional (dimension $4$ in our case).
The projection of $W$ on $\mcK_0$ with respect to $g_0$ reads $\sum_{j} \langle e_j,W\rangle_{g_0} e_j$, 
where $\big(e_j\big)_{1\leq j \leq 4}$ is an orthonormal basis of $\mcK_0$ for the $g_0$-norm, whereas it reads
$\sum_{j} \langle e_j^{g'},W\rangle_{g'} e_j^{g'}$ when the projection is done with respect to $g'$, 
where $\big(e_j^{g'}\big)_{1\leq j \leq 4}$ is the orthonormal basis of $\mcK_0$ with respect to the $g'$-norm
obtained as the Gram-Schmidt basis from $(e_j)$ in the $g'$-norm. 
We wish to show that we have the convergence $e_j^{g'}$ towards $e_j$ as $g' \rightarrow g_0$ for all $j$.
To do so, consider the map
$$\Theta : \left(g',u_1,\dots,u_4\right) \mapsto \left(\langle u_j,u_k \rangle_{g'}\right)_{1 \leq k,l \leq 4}\ ,$$
defined and smooth for $g'$ in a neighborhood of $g_0$ and matrices $[u_1,\ldots,u_4]$ being upper triangular 
with positive diagonal coefficients. 
The differential of this map restricted to the $u_i$-variables writes 
$$D_2 \Theta (g_0,e_1,\ldots,e_4) : 
(u_1,\ldots,u_4) \mapsto \left(\langle e_j,u_k \rangle_{g_0} + \langle e_k,u_j \rangle_{g_0}\right)_{1 \leq k,l \leq 4}\ ,$$
in other words, it maps a matrix $U = [u_1,\ldots,u_4]$ to its symmetric sum $U + U^{T}$, so that its kernel, when restricted to
the space of upper triangular matrices, is trivial.
One can therefore apply the implicit function Theorem to $\Theta$, and get the existence of a neighborhood $\mcV_0$ of
$(e_1,\ldots,e_4)$ in $(\mcK_0)^4$ and a neighborhood $\mcU_{g_0}$ of $g_0$ such that, for $(u_1,\ldots,u_4)$ and $g'$
in these neighborhoods, one has the equivalence 
$$\Theta (g',u_1,\ldots,u_4) = I \Longleftrightarrow \forall j\ u_j = u_j(g')\ ,$$
with $g' \mapsto u_j(g')$ being smooth functions of $g'$. But the left-hand side of the above equivalence implies that 
$(u_1,\ldots,u_4)$ is an orthonormal basis of $\mcK_0$ for the $g'$-norm, and it is written as a upper-triangular system 
with positive diagonal coefficients in the $(e_1,\ldots,e_4)$-basis: in other words, $u_j(g')$ coincide with $e_j^{g'}$ forall
all $1 \leq j \leq 4$, and the $e_j^{g'}$ converge to $e_j$ as $g' \rightarrow g_0$, as desired.  
Then, the fact that the convergence~(\ref{convproj}) is (locally) uniform with respect to $W$ comes easily from this fact.

\medskip

For the expression $IV$, we note that the projection operator $\Pi_{\mathcal K_0 ^{\perp g_0}}$ is continuous (hence bounded)
in the $g_0$-norm, so one has
$$IV \leq C \|e^{-\frac1x}\big[\Phi\big(G' + F^2 e^{\frac1x}P^* z \big) - \Phi\big(G_0 + F^2 e^{\frac1x}P_0^* z \big)\big] \|_{g_0}\ ,$$
which converges to $0$ as $(i,m,\vec a)$ tends to $(\infty,m_0,\vec 0)$ uniformly in $z$ in some ball $B_{E(g_0)}(0,r_0)$, since 
the map $G' \mapsto e^{-\frac1x}\Phi(G')$ is three times differentiable in a neighborhood of $G_0$, 
hence its second derivative is Lipschitz in a neighborhood of $G_0$ (in suitable $g_0$-norms).

\medskip

To complete the proof of the proposition, we notice that the norms computed with respect to $g'$ and to $g_0$ are equivalent
for all $(m,\vec a)$ in a small neighborhood of $(m_0,\vec 0)$ and $i$ large enough, so that we still obtain the uniform convergence of 
$$\frac{\|Q_{i,m,\vec a}(y) - Q_{i,m,\vec a}(x)\|_{F(g')}}{\|y-x \|_{E(g')}\big(\|x\|_{E(g')} + \|y\|_{E(g')}\big)}$$
towards 
$$\frac{\|Q_{\infty,m_0,\vec 0}(y) - Q_{\infty,m_0,\vec 0}(x)\|_{F(g_0)}}{\|y-x \|_{E(g_0)}\big(\|x\|_{E(g_0)} + \|y\|_{E(g_0)}\big)}$$
as $(i,m,\vec a) \rightarrow (\infty,m_0,\vec 0)$, uniformly with respect to $x$ and $y$ in some ball of radius $r_0$ 
for the $E(g')$ norm, with $x \neq y$, for some $r_0 > 0$. 
Hence, there exists a constant $r_1$ in $(0,r_0)$ such that the inequality announced holds 
for all $x,y$ in $B_{E_{g_0}}(0,r_1)$ and for all $i \geq r_1^{-1}$. 

\qed

\bigskip

The above proposition gives us some constants $q > 0$, $r_0 > 0$, $i_0 \in \mathbb{N}$ and $\delta_0 >0$
such that the second requirement of the Lemma~\ref{ptfixe} above is fulfilled independently of $(i,m,\vec a)$
with $i \geq i_0$ and $|(m,\vec a) - (m_0,\vec 0)| < \delta_0$, so that for all $r < \min \left(\frac{1}{2qC},r_0\right)$,
if we further have $\|\Psi_{i,m,\vec a}(0)\|_{F(g')} \leq \frac{r}{2C}$.
Such $r$ do exist, provided one, if necessary, increases $q$ and $i_0$, and decreases $r_0, \delta_0 >0$, in order to obtain
smaller bounds on the norm of $\Psi_{i,m,\vec a}(0)$. 

In other words, for all such $(i,m,\vec a)$, there exists a (locally unique) pair of the form $(\delta g',\delta k')$, 
such that the equality
\bel{tildeg}\Pi_{\mathcal K _0 ^{\bot g'}}\ e^{-1/x} \Phi\left((g',k') + (\delta g',\delta k')\right) = 0\;\ee
holds.

\section{Solving the constraint equation}
\label{sec_solve}

\subsection{Projection on the cokernel}

At the last section~\ref{sec_deform}, we proved the existence of a pair $(\tilde g,\tilde k)$ of initial data 
which matches with the original pair
$(g,k)$ in a complement of the asymptotic region, and which coincides with the Kerr-Kottler-de Sitter initial data
$(\mathring g_{m,\vec a},\mathring k_{m,\vec a})$ far enough in the asymptotic region, with an intermediate gluing zone 
where we have performed a perturbation of the initial interpolation $(g',k')$, so that the perturbed data
$(\tilde g,\tilde k)$ pointwise satisfy $\Pi_{\mathcal{K}_0^{\bot g'}}e^{-1/x}\Phi(\tilde g,\tilde k) = 0$.

Note that if the kernel $\mathcal K_0$ were trivial, then the initial data $(\tilde g,\tilde k)$ would be solution
of the constraint equations $\Phi(\tilde g,\tilde k) = 0$ and we would conclude.
However, the kernel $\mathcal K_0$ is not trivial, so let us consider the maps
\bel{projK_0}
F_i : \begin{array}{ll} B((m_0,\vec 0),\delta_0) \ \subset \ \R ^4 \ \longrightarrow \ \ \R^4
\\ \ \ \ \ \ \ (m,\vec a) \ \ \ \ \ \ \ \ \  \mapsto \ \ \  \left(\int_{\Omega} \langle\Phi(\tilde g,\tilde k),V_{\mu}\rangle d\mu_{g_0}\right)_
{\mu = 0\ldots 3}
 \end{array}\;,
\ee
which are the $L^2 _{e^{1/2x}}(g_0)$ projections on $\mathcal K_0$ of $e^{-1/x}\Phi(\tilde g,\tilde k)$.

\medskip

The maps $F_i$ are all well defined for all $i \geq i_0$ on the ball $B((m_0,\vec 0),\delta_0)$
of $\R^4$, where $i_0 \in \N$ and $\delta_0 > 0$ have been obtained in the previous section~\ref{sec_deform}.

\medskip

We wish to prove the existence, for every integer $i$ large enough, of a couple $(m,\vec a)$ close to $(m_0,\vec 0)$
and such that $F_i(m,\vec a) = 0$;
this would provide us with data $(\tilde g,\tilde k)$ whose projection on $\mathcal K_0$ is zero. 
Combining this with~(\ref{tildeg}), we would conclude that $\Phi(\tilde g,\tilde k) = 0$.
To do so, we will use for the $F_i$'s the following result, which is a consequence
of the Brouwer fixed point Theorem (see~\cite{CD03}, Lemma 3.18):
\begin{prop}\label{Glambda=0}
Let $G: U \rightarrow V$ be a homeomorphism between two open sets $U$ and $V$ of $\R^d$. Let 
$\{G_{\lambda}\}_{\lambda \in \R}$ be a family of continuous maps
$U \rightarrow \R^d$, which uniformly converge to $G$ as $\lambda$ tends to $+\infty$. Then, for all $y \in V$,
there exists $\lambda_0$  such that the condition $\lambda \geq \lambda_0$ implies that the equation $G_{\lambda}(x) = y$
admits a unique solution $x_{\lambda}$.
\end{prop}

The remainder of this section is devoted to the computation of the $F_i$'s, in order to check that we can apply the proposition, 
with the parameter $i$ having the role of $\lambda$.

\subsection{Global charges}

We start this paragraph by deriving the \emph{global charges} of initial data asymptotic to a model, 
following the notations of \cite{thesebenoit}.
If $(g_0,k_0)$ is the background initial data and $(g_1,k_1)$ is ``close'' to $(g_0,k_0)$, 
then, let $e := (g_1 - g_0,k_1 - k_0)$ be the difference between them and let $V \in \mathcal K_0$,
one can write
$$ \int_{\Omega} \langle \Phi (g_1,k_1) , V \rangle d\mu_{g_0} - \int_{\Omega} \langle \Phi (g_0,k_0) , V \rangle d\mu_{g_0} =
 \int_{\Omega} \langle D\Phi_{(g_0,k_0)}(e), V \rangle d\mu_{g_0} + Q(V,e)\;,$$
where $Q(V,e) := \int_{\Omega}Q(e) V d\mu_{g_0}$ and where $Q(e)$ is obtained as
\bel{Q(e)}
Q(e) = \Phi(g_1,k_1) - \Phi(g_0,k_0) - D\Phi(g_0,k_0)(e)\;.
\ee
Since $D\Phi_{(g_0,k_0)}^*$ is the $L^2$ formal adjoint of $D\Phi_{(g_0,k_0)}$, one has
\bel{divU}\langle D\Phi_{(g_0,k_0)}(e), V \rangle = \langle D\Phi_{(g_0,k_0)}^* V, e \rangle + \mathrm{div} \mathbb U (V,e)\;,
\ee
and since $\Phi (g_0,k_0) =0$ and $V \in \mathcal K_0$, one eventually has
\bel{charge}\int_{\Omega} \langle \Phi (g_1,k_1) , V \rangle d\mu_{g_0} = \oint _{\partial \Omega} \mathbb U(V,e)(\nu) dS + Q(V,e)\;,\ee
where $\nu$ is the outwards unit normal vector of $\partial \Omega$ and $dS$ the induced measure on it, with respect to $g_0$.
The \emph{global charge associated to $V$} is the boundary integral term $\partial \Omega$ de $\Omega$ in the formula~(\ref{charge}).

For $V = (f,\alpha) \in \mathcal K_0$, one has the explicit formula (see~\cite{thesebenoit})
\bel{U(V,e)}
\mathbb U(V,e) = f\left(\mathrm{div}\;h - d(\mathrm{tr}\;h)\right) - \iota_{\nabla f}h + (\mathrm{tr}\;h)df +
2\left(\iota_{\alpha}l - (\mathrm{tr}\;l)\alpha\right)\;.
\ee
In particular, one notices that this expression is linear with respect to $V = (f,\alpha)$.

\subsection{Computation in the Schwarzschild-de Sitter case}

In the present case, $g_0 = b_{m_0}$, $k_0 = 0$ (SdS initial date with mass $m_0$), $\Omega = (\delta,T - \delta) \times \mathbb S^2$.
Let us first take $g_1 = \mathring g_{m,\vec a}$,
$k_1 = \mathring k_{m,\vec a}$ (KdS initial data with parameters $m,\vec a$), we note $h = g_1 - g_0$,
$l = k_1 - k_0$, in other words $e = (h,l)$.

For $V = V_0 = (f,0)$, we compute $\mathbb U(V_0,e)$
in the $(r,\theta,\varphi)$ coordinate system, where $g_0$ and $g_1$
are diagonal:
$$\mathbb U(V_0,e)(\nu) = f^2 \left(g_0^{rr} \partial_r h_{rr} - \partial_r \left(\tr_0 h\right)\right) +
 f\left(-f^2 (\partial_r f) h_{rr} + (\tr_0 h)\partial_r f\right)\;,
$$
or:
$$\mathbb U(V_0,e)(\nu) = f^2 \left(- \partial_r(f^2)h_{rr} - \partial_r \left(g_0^{\theta\theta}h_{\theta\theta}
+ g_0^{\varphi\varphi}h_{\varphi\varphi}\right)\right) + f\left(g_0^{\theta\theta}h_{\theta\theta}
+ g_0^{\varphi\varphi}h_{\varphi\varphi}\right)\partial_r f\;.
$$
We find that $h_{rr} = \frac{2 (m - m_0)}{r f^4} + O(|(m-m_0,\vec a)|^2)$
and $g_0^{\theta\theta}h_{\theta\theta} + g_0^{\varphi\varphi}h_{\varphi\varphi} = O(a^2)$, as long as its derivative with respect to $r$.

We therefore obtain
\bel{U(V_0,e)}\mathbb U (V_0, e)(\nu) =
\frac{-2\partial_r (f^2)}{r f^2}(m - m_0) + O(|(m-m_0,\vec a)|^2)\;.\ee
The computations above are valid provided that we choose $r$ such that $f(r) \neq 0$. 
The integration on the slice $\{r\}\times \mathbb S^2$ yields

$$\int_{\{r\}\times \mathbb S^2}\mathbb U(V_0,e)(\nu)dS = -\frac{8\pi r \partial_r(f^2)}{f^2}(m-m_0) + O(|(m-m_0,\vec a)|^2)\;.$$

\medskip

Then, we choose spherical coordinates such that $\vec a = a \partial_z$.
In thus have $V_3 = (0,\partial_{\varphi})$, and can compute
$$\mathbb U(V_3,e)(\nu) = 2 f \mathring k_{r\varphi}\;, $$
where $\mathring k$ refers to $\mathring k_{m,\vec a}$. Hence, from~(\ref{krphi}), one has
\bel{U(V_3,e)}
\mathbb U(V_3,e)(\nu) = -\frac{6 m_0 \sin^2 \theta}{r^2} a +
 O(|(m-m_0,\vec a)|^2)\;,
\ee
and thus
$$ \int_{\{r\}\times \mathbb S^2}\mathbb U(V_3,e)(\nu)dS
= - 16 \pi m_0 a + O(|(m-m_0,\vec a)|^2)\;.$$
On the other hand, the elements $V_1 = (0,X_1)$ and $V_2 = (0,X_2)$ of $\mathcal K_0$ can be written in coordinates $r,\theta,\phi$ as:
$$X_1 = - \sin \varphi \partial_{\theta} - \cot \theta \cos \varphi \partial_{\varphi}\;,$$
and
$$X_2 = - \cos \varphi \partial_{\theta} + \cot \theta \sin \varphi \partial_{\varphi}\;.$$
When one integrates over $2\pi$ period in $\varphi$, it yields
$$  \int_{\{r\}\times \mathbb S^2}\mathbb U(V_i,e)(\nu)dS = 0\;,$$
for $i = 1,2$.

The above calculations have been carried out in the spherical coordinates $(r,\theta,\varphi)$ with respect to cartesian coordinates
$(x,y,z)$ (namely $\theta$ measures the colatitude, $\varphi$ the radius...) such that the parameter $\vec a$ is directed
along $\partial_z$, $\vec a = a \partial_z$.

With the notations $\vec V = (V_1,V_2,V_3)$, we have therefore obtained
$$\int_{\{r\}\times \mathbb S^2}\mathbb U(\vec V,e)(\nu)dS =
 - 16 \pi m_0 \vec a + O(|(m-m_0,\vec a)|^2)\;.$$

Since this is true for an arbitrary choice of coordinates $x,y,z$, one has, for all vector $\vec a$ in $\R^3$ with a small enough norm,
the general result,
$$ \int_{\{r\}\times \mathbb S^2}\mathbb U(V_i,e)(\nu)dS =
 - 16 \pi m_0 a_i + O(|(m-m_0,\vec a)|^2)$$
for all $i \in \{1,2,3\}$.

\subsection{Charges of perturbed data}

Let us now consider $g_1 = \tilde g$, $k_1 = \tilde k$ (depending on $i,m,\vec a$). Since $(g_1,k_1)$ coincides with
$(\mathring g_{m,\vec a},\mathring k_{m,\vec a})$ at $r = r(y = T - \delta)$ to any order, 
the integrals
$$\int_{\{r\}\times \mathbb S^2}\mathbb U(V_\mu,e)(\nu)dS\;$$
still have the same values as above for $\mu \in \{0,\cdots,3\}$ for this value of $r$. 
In fact, these expressions do not depend on $i$.
On the other hand, $(g_1,k_1)$ coincides to any order with the original initial data $(g_i,k_i)$ en $r = r(y= \delta)$,
so that for this value of $r$, the integrals
$$\int_{\{r\}\times \mathbb S^2}\mathbb U(V_\mu,e)(\nu)dS$$
do not depend on $(m,\vec a)$ and converge to $0$ when the parameter $i$ tends to $+\infty$, since
$(g_i,k_i) \rightarrow (g_0,0)$.

One can summarizes by writing
\bel{F_i(m,a)}
F_i(m,\vec a) = -8\pi \left(\frac{r^2 \partial_r(f^2)}{f^2}(m-m_0), 2 m_0 \vec a\right) +  O(|(m-m_0,\vec a)|^2) + o(1)
+ (Q(V_{\mu},e))_{\mu=0\ldots 3}\;,
\ee
where $r = r(y= T - \delta)$.
The term $o(1)$ above does not depend on $(m,\vec a)$, hence it does uniformly converge to $0$ as
$i$ tends to infinity, and meanwhile, $O(|(m-m_0,\vec a)|^2)$ does not depend on $i$.

\bigskip

We now have to evaluate the terms $Q(V,e)$, for $V = V_{\mu}$, $\mu=0\ldots 3$. 
More precisely, we would like these functions to uniformly converge with respect the parameters $(m,\vec a)$ 
as $i \rightarrow +\infty$, and the limit to read
$o(|(m-m_0,\vec a)|)$.

To do so, one considers the initial data $ (g'_{\infty},k'_{\infty})$ obtained by making the interpolation between
the SdS initial data $(g_0,0)$, and the KdS initial data $(\mathring g_{m,\vec a},\mathring k_{m,\vec a})$, and we still
use the cut-off function $\chi$ for this interpolation. 
We therefore have 

$$ (g'_{\infty},k'_{\infty}) := (1-\chi) (g_0,0) + \chi (\mathring g_{m,\vec a},\mathring k_{m,\vec a})\;.$$
Notations with indices $\infty$ indicate that this interpolation obtained as a limit as $i \rightarrow +\infty$
of the interpolation $(g',k')$ between $(g_i,k_i)$ and $(\mathring g_{m,\vec a},\mathring k_{m,\vec a})$ given by 
the formula~(\ref{interpolation}).
Following exactly the same procedure as in section~\ref{sec_deform}, for each $(m,\vec a) \in B((m_0,\vec 0),\delta_0)$, 
one can likewise show the existence of a couple
 $(\delta g_0,\delta k_0)$ such that
$$\Pi_{\mathcal K _0 ^{\bot g'_{\infty}}}\ e^{-2s/x} \Phi(\tilde g_{\infty},\tilde k_{\infty}) = 0\;,$$
where $\tilde g_{\infty} = g'_{\infty} + \delta g_0$, $\tilde k_{\infty} = k'_{\infty} + \delta k_0$.
Recall that these quantities do not depend on $i$, because $\sigma$ has been chosen so that the initial data
$(\mathring g_{m,\vec a,i,\sigma},\mathring k_{m,\vec a,i,\sigma})$ do not depend on $i$.

Let us now consider~(\ref{charge}) with $(g_1,k_1) = (\tilde g_{\infty},\tilde k _{\infty})$.
Denoting $e_0 := (\tilde g_{\infty} - g_0,\tilde k _{\infty})$, for $V \in \mcK_0$, one has
$$\int_{\Omega} \Phi(\tilde g_{\infty},\tilde k _{\infty})V d\mu_{g_0} = \oint_{\partial \Omega} \mathbb U(V,e_0)(\nu)dS
+ Q(V,e_0)\;.
$$
As for the couple $(\tilde g,\tilde k)$ above, one has
$$\Big(\oint_{\partial \Omega} \mathbb U(V_{\mu},e_0)(\nu)dS\Big)_{\mu = 0\ldots 3} = -8\pi \left(\frac{r^2 \partial_r(f^2)}{f^2}(m-m_0), 2 m_0 \vec a\right) +
O(|(m-m_0,\vec a)|^2)\;.$$
There is no more dependence on $i$. On the other hand, since the error term $Q(V,e_0)$ does not depend on $i$, one 
wishes to show that it is limit of $Q(V,e)$ as $i \rightarrow +\infty$, the convergence being uniform 
with respect to $(m,\vec a) \in B((m_0,\vec 0),\delta_0)$.
To that end, we notice that
$$e = e_0 + (\tilde g - \tilde g_{\infty},\tilde k - \tilde k_{\infty})\;.$$
we have to estimate the second term at the right-hand side of this equality.
The conclusion will then follow from:
\bel{Q(e)-Q(e_0)}
Q(V,e) = Q(V,e_0) + \int_{\Omega} V \left[Q(e) - Q(e_0)\right]d\mu_{g_0}\;.
\ee
The writing of this splitting highlights the term 
$Q(V,e_0)$ which does not depend on $i$ on the one hand;
on the other hand, from the definition of $Q$ (see~(\ref{Q(e)})), and from the regularity of
$\Phi$, one can write the pointwise inequality:
$$|Q(e) - Q(e_0)| \leq q |e - e_0| \big(|e| + |e_0|\big)
$$
for all values of $(i,m,\vec a)$ such that $i \geq i_0$,
$|(m-m_0,\vec a)| < \delta_0$.

We compute $e - e_0$, which yields
$$(\tilde g - \tilde g_{\infty},\tilde k - \tilde k_{\infty}) =
(\delta g,\delta k) - (\delta g_0,\delta k_0) + (1-\chi) (g_i - g_0,k_i)\;.$$
For the sake of clarity, we denote $G_1$ for the couple $(g_1,k_1)$.
With respect to these new notations, one has:
$$e - e_0 = \tilde G - \tilde G_{\infty} = \delta G - \delta G_0
+ (1- \chi)(G_i - G_0)\;.$$
Now, using the property of $Q$,
one has
$$|Q(e) - Q(e_0)| \leq q_0 |\delta G - \delta G_0
+ (1- \chi)(G_i - G_0)| \big(|e| + |e_0|\big)\;,
$$
with:
$$
|e| + |e_0| \leq 2|e_0| + |\delta G - \delta G_0| + |(1 - \chi)(G_i - G_0)| \;.
$$
In particular, the factor $|e| + |e_0|$ is bounded independently of $i,m,\vec a$ with $i \geq i_0$ and $|(m-m_0,\vec a)| < \delta_0$.
On the other hand, the term $|(1-\chi)(G_i - G_0)|$ does not depend on $(m,\vec a)$ and tends to $0$ as $i$ tends to $+\infty$.
In order to determine the convergence of $|\delta G - \delta G_0|$, one can simply notice that the map
$G' \mapsto \delta G$ is uniformly continuous. In fact, as it appears when considering the parameters $i,m,\vec a$, the sequence
$i \mapsto \delta G_{i,m,\vec a}$ converges to $\delta G_{\infty,m,\vec a}$, uniformly with respect to $(m,\vec a)$ with 
$|(m-m_0,\vec a)| < \delta_0$.
Indeed, this property arises from the application of Picard's fixed point Theorem with parameter, that we use to solve the projected
problem~(\ref{tildeg}), see for example the Proposition G.1 of~\cite{CD03}.

\medskip

To summarize, we have bounded $|Q(e) - Q(e_0)|$ by a term converging to $0$ as $i$ tends to the infinity uniformly with respect to 
$(m,\vec a)$ such that $|(m-m_0,\vec a)| < \delta_0$.
The formula~(\ref{Q(e)-Q(e_0)}) then allows us to assert that, for any $V \in \mcK_0$, $Q(V,e)$ converges to $Q(V,e_0)$ as
$i \rightarrow +\infty$, uniformly in $(m,\vec a)$ such that 
$(m,\vec a) \in B((m_0,\vec 0),\delta_0)$.

%
%
%
%
%
%
%
%
%

Thus, one can use Proposition~\ref{Glambda=0} for the family of maps $(F_i)_{i \geq i_0}$,
since it converges uniformly to a map $F$, which reads:
$$F(m,\vec a) =  -8\pi \left(\frac{r^2 \partial_r(f^2)}{f^2}(m-m_0), 2 m_0 \vec a\right)
+  O(|(m-m_0,\vec a)|^2)\;.$$
Hence, taking a smaller $\delta_0$ if necessary, with $U = B((m_0,\vec 0),\delta_0)$ and $V = F(U)$, the map $F$ 
is a homeomorphism between $U$ and $V$ with $0 \in V$ since $F(m_0,\vec 0) = 0$.
Therefore, by increasing $i_0$ if necessary, one obtains that, for all $i \geq i_0$,
there exists a unique $(m_i,\vec a _i) \in U$ such that $F_i(m_i,\vec a _i) = 0$.
The couple $(\tilde g,\tilde k)$ corresponding to the parameters $(m_i,\vec a_i)$ 
is solution of the constraint equations, which finishes the proof of the Theorem \ref{thmrecoll}.

We observe that the initial date then obtained are nontrivial in general (meaning that they do coincide with KdS initial data in the
asymptotic region, but not globally), since they coincide with the original initial data $(g,k)$
which asymptote to $(g_0,k_0) = (b_{m_0},0)$.

\begin{engrk}
It appears from the last two sections that the assumptions of the Theorem \ref{thmrecoll} may not be optimal, since
initial data $(g,k)$ such that $(g-g_0,k-k_0)$ has a small enough $g_0$-norm in the asymptotic region can be treated as above. 
\end{engrk}

\begin{engrk}
One could derive such gluing result starting from an initial data set $G$ asymptotic to Schwarzschild-de Sitter,
such that $\mu \geq |\mathbf{J}|_{g_0}$, where $\mu$ is the function equal to the first line of the constraint operator defined 
in~(\ref{contraintes}), 
while $\mathbf J$ is the one-form equal to the second line of it.
In the spirit of the work of Delay in~\cite{Del11} (see in particular the section 4 and Theorem 4.1 there), 
one could construct a gluing between $G$ and some $\mathring G_{m,a}$ as above, 
imposing only that the inequality $\mu \geq |\mathbf J|_{g_0}$ is everywhere preserved. 
\end{engrk}

\section{Gluing constructions for asymptotically Kerr-de Sitter initial data}

One can reproduce step by step the above work, for asymptotically Kerr-de Sitter initial data.
We consider initial data $G = (g,k)$ which asymptote to the model initial data 
$G_0 = (g_0,k_0) := (\mathring{g}_{m,a},\mathring{k}_{m,a})$ for some $m$ and $a$ 
admissible and non-zero, that is to say, such that the four roots $r_1,\ldots,r_4$ defined in section~\ref{sec_kds} 
exist and are distinct.
As seen in section~\ref{sec_kds}, these initial data are also periodic in $y$, with period $T = T(m,a)$ (and we still 
impose the condition $\Lambda =3$ as before).

In this case, the co-kernel $\mathcal K_0$ of the linearized constraint operator evaluated at $G_0$ contains at least
the Killing initial data of the Killing fields $\partial_t$ and
$\partial_{\varphi}$ of the Kerr-de Sitter space-time in Boyer-Linquist coordinates. 
It is a widely known fact that there is no other independant Killing field. 
The proof of this for Kerr ($\Lambda = 0$) space-times can be found in~\cite{One95},
Theorem 3.8.8, together with a complete determination of the group of isometries.

Hence, $\mathcal K_0$ is spanned by the corresponding elements $V_0$ and $V_1$, with
$V_0 = (f_0,Z_0)$, where $f_0 = \sqrt{\frac{-\lambda}{g_{\varphi \varphi}}}$, and
$Z_0 = \frac{\bar g_{t\varphi}}{g_{\varphi \varphi}}\partial_{\varphi}$,
and with $V_1 = (f_1,Z_1)$, where $f_1 = 0$ and $Z_1 = \partial_{\varphi}$.
$\lambda$ refers to the (negative) quantity $\bar g_{tt} \bar g_{\varphi \varphi} - \bar g_{t\varphi}^2$.

On the other hand, the candidate family of initial data used to carry out the gluing will be the Kerr-de Sitter family of initial data 
with a fixed direction given by $\vec a = a \partial_z$ for the angular momentum.
This family has therefore two parameters $m'$ and $a'$.

For each admissible value of $(m',a')$, and for $i \in \mathbb N$, 
one can again define an interpolation $G'_{i,m,\vec a} := (g'_{i,m',a'}, k'_{i,m',a'})$ between 
$G$ and $\mathring G_{m',a'}$ with a cut-off function $\chi$ translated such that its support
is included in a (relatively compact) region $\Omega_{i,T,T'}$, with boundary and a defining function $x$, 
whose definition is similar to the one given in section~\ref{sec_deform}, where 
$T = T(m,a)$ and $T' = T(m',a')$ are the respective periods of both initial data. 
See again the Figure~\ref{gluing}, with now $(g_0,k_0)= (\mathring{g}_{m,a},\mathring{k}_{m,a})$.

Then, for each $(m',a')$ close enough to $(m,a)$ and for all $i$ big enough,
one can solve the projected problem in $\delta G'$:
$$\Pi_{\mcK_0 ^{\perp g'}}e^{-\frac1x} \Phi\big(G'_{i,m',a'} + \delta G'\big) = 0\ ,$$
and, repeating the arguments of section~\ref{sec_deform}, find a locally unique solution 
$\delta G'_{i,m',a'}$, small in $g_0$-norm, of the above equation, and we again note 
$\tilde G_{i,m',a'} = G'_{i,m',a'} + \delta G'_{i,m',a'}$.

For the projection on $\mcK_0$, one needs to compute the boundary integrals likewise section~\ref{sec_solve}, with now the background 
$G_0 = (g_0,k_0)$ being that of Kerr-de Sitter with parameters $m$ and $a$. In particular, we need to evaluate 
the integrals 
$$ \int _{\{r\}\times \mathbb S^2} \mathbb U(V,e)(\nu) dS\ ,$$
where $V \in \mcK_0$ and $e = (g_1 - g_0, k_1 - k_0) = (h,l)$, where $(g_1,k_1)$ are the KdS initial data with parameters $(m',a')$ close to
$(m,a)$.
We recall the general expression of the charge integrand $\mathbb U(V,e)$ given in~\cite{thesebenoit}, for $V = (f,\alpha)$:
\bea \mathbb U(V,e) & = & \nonumber{f\left(\mathrm{div}\;h - d(\mathrm{tr}\;h)\right) - \iota_{\nabla f}h + (\mathrm{tr}\;h)df +
2\left(\iota_{\alpha}l -(\mathrm{tr}\;l)\alpha\right)}\\
& & + (\mathrm{tr}\;h)\iota_{\alpha} k_0 + g_0(k_0,h) \alpha  
-2 \iota_{\alpha} (h \circ k_0)\ .\eea
The traces and divergences above are computed with respect to $g_0$.
For such $G_1 = (g_1,k_1)$, we obtain expressions of the form 
$$\Big(\int _{\{r\}\times \mathbb S^2} \mathbb U(V_{\mu},e)(\nu) dS\Big)_{\mu = 0,1} = \big(u(r)(m'-m),A(a'-a) \big) + O(|(m'-m,a'-a)|^2)\ ,
$$
for some function $u$ independent of $m',a'$ and some non-zero constant $A$.

Explicitely, with the help of the {\sc Mathematica} program, one finds formulas, which we only give here for small values of $a$ 
to avoid very long expressions:

\begin{itemize}
 \item For $V=V_0$, and for $a'= a$, one has

$$\int _{\{r\}\times \mathbb S^2} \mathbb U(V_0,e)(\nu) dS = 
\left(-\frac{16\pi \left(-m+r^3\right)}{r^2 \left(2 m-r+r^3\right)}+O[a]^2\right)(m'-m)
 + O((m'-m)^2)\ ,
$$
whereas for $m'=m$,
$$
\int _{\{r\}\times \mathbb S^2} \mathbb U(V_0,e)(\nu) dS = 
\left(\frac{\pi P_m(r) a}{6 r^5 \left(2 m-r+r^3\right)}+O[a]^2\right)(a'-a) + O((a'-a)^2)\ ,
$$
where $P_m(r)$ is the polynomial given by

\beaa
P_m(r) = 4 m^3 (-160+27 \pi )+2 m^2 r \left(128-27 \pi +6 (-80+9 \pi ) r^2\right)\\
+m r^2 \left(192+(64-27 \pi ) r^2+3 (-64+9 \pi ) r^4\right)+64 r^3 \left(-2+6 r^2-5 r^4+r^6\right)\ .
\eeaa

\item For $V = V_1 = (0,\partial_{\varphi})$, and for $a'=a$, one has

$$\int _{\{r\}\times \mathbb S^2} \mathbb U(V_1,e)(\nu) dS = 
\left(-\frac{9 \pi^2  \left(-1+r^2\right) a}{2 r \left(2 m-r+r^3\right)}+O[a]^2\right)(m'-m) + O((m'-m)^2)\ ,
$$
whereas for $m'=m$, one has
$$\int _{\{r\}\times \mathbb S^2} \mathbb U(V_1,e)(\nu) dS = 
\left(-\frac{9 m \pi^2 }{2 r^2} + O[a]^2\right)(a'-a) + O((a'-a)^2)
 \ .
$$
\end{itemize}

Hence, for all $a$ small enough and for any admissible $m$, all expressions above have non trivial coefficients in front of the order one terms in $(m'-m)$
or $(a'-a)$, for $r$ taking values on the complement of a discrete subset of the interval $(r_3,r_4)$, as desired.

Then, for $(g_1,k_1) = \tilde G_{i,m',a'}$, one has again the same expression for the boundary integrals plus an error term
of the form $o(1)$ as $i \rightarrow \infty$, whereas the remainder $Q(V,e)$ defined as in the previous section is dominated by a term
$O(|(m'-m,a'-a)|^2) + o(1)$ as $i \rightarrow \infty$.   

Thus, one gets the similar conclusion as before, applying the homeomorphism result stated in the Proposition~\ref{Glambda=0} to the maps
$F_i$ which give the projection on $\mcK_0$, as defined in~\ref{F_i(m,a)}, since the map obtained as the limit as $i \rightarrow \infty$ 
will be itself a local homeomorphism, sending $(m,a)$ to $0$.

We therefore obtain that for every big enough $i$, there exists a (locally unique) pair $(m',a')$ such that the projection on $\mcK_0$ 
vanishes, namely such that $\tilde G_{i,m',a'}$ is the desired solution of the constraint equations.

%
%
%

\bibliographystyle{amsplain}
\bibliography{these2}

\end{document}